\documentclass[letterpaper, aps, prd, twocolumn, superscriptaddress, showpacs, nofootinbib]{revtex4}
\pdfoutput=1

\usepackage{graphicx}
\usepackage{bm}
\usepackage{dcolumn}
\usepackage{color}
\usepackage[sort&compress]{natbib}
\usepackage{url}
\usepackage{subfigure}
\usepackage{longtable}
\usepackage{amssymb}
\usepackage{amsmath}
\usepackage{amsfonts}

\def\apj{Astrophys.\ J.}
\def\apjl{Astrophys.\ J. Lett.}

\def\aap{Astron.\ Astrophys.}

\def\mnras{Mon.\ Not.\ Roy.\ Astron.\ Soc.}

\def\prd{Phys.\ Rev.\ D}

\def\cqg{Class.\ Quantum Grav.}

\begin{document}

\title{Reconstructing Gravitational Wave Core-Collapse Supernova Signals with Dynamic Time Warping}

\newcommand*{\UoMelbsp}{OzGrav, School of Physics, University of Melbourne, Parkville, VIC 3010, Australia.}
\affiliation{\UoMelbsp}
\newcommand*{\UoMelbs}{OzGrav, University of Melbourne, Parkville, VIC 3010, Australia.}
\affiliation{\UoMelbs}
\newcommand*{\Swinburne}{OzGrav, Swinburne University of Technology, Hawthorn, VIC 3122, Australia.}
\affiliation{\Swinburne}

\author{Sofia Suvorova} \affiliation{\UoMelbs}
\author{Jade Powell}  \affiliation{\Swinburne}
\author{Andrew Melatos}	\affiliation{\UoMelbsp}

\date{\today}

\begin{abstract} 
Core-collapse supernovae (CCSNe) are a potential source for ground-based gravitational wave detectors, as their predicted emission peaks in the detectors' frequency band. Typical searches for gravitational wave bursts reconstruct signals using wavelets. However, as CCSN signals contain multiple complex features in the time-frequency domain, these techniques often struggle to reconstruct the entire signal. An alternative method developed in recent years involves applying principal component analysis (PCA) to a set of simulated CCSN models. This technique enables model selection between astrophysical CCSN models as well as waveform reconstruction. However, PCA faces its own difficulties, such as being unable to reconstruct signals longer than the simulations; many CCSN simulations are stopped before the emission peaks due to insufficient computational resources. In this study, we show how combining PCA with dynamic time warping (DTW) improves the reconstruction of CCSN gravitational wave signals in Gaussian noise characteristic of Advanced LIGO at design sensitivity. For the waveforms used in this analysis, we find that the number of PCs needed to represent 90\% of the data is reduced from nine to four by applying DTW, and that the match between the original and reconstructed waveforms improves for signal-to-noise ratios in the range $[0,50]$.
\end{abstract}

\maketitle

\section{Introduction}
\label{sec:intro}

A potential source for ground based gravitational wave interferometers such as Advanced LIGO \cite{aLIGO} and Advanced Virgo \cite{AdVirgo} is a burst signal from a nearby core-collapse supernova (CCSN) \cite{2016PhRvD..93d2002G, 2016PhRvD..94j2001A}. Gravitational waves from a CCSN are emitted from deep inside the core, from the region where electromagnetic radiation does not escape, opening a new window on the core collapse process. The most recent CCSN gravitational waveforms are currently only partially modelled due to complicated physics and computationally expensive simulations; see \cite{2017arXiv170304633M} for a recent review. Most of the gravitational wave emission is expected to be emitted within the first $\sim 1\,{\rm s}$ of the explosion, due to convection and the standing accretion shock instability (SASI) \cite{2003ApJ...584..971B, andresen:16, mueller:e12, kuroda:16, morozova_18, 2017arXiv170107325Y}.  

The waveforms contain stochastic components which rule out detection techniques such as matched filtering \cite{1999PhRvD..60b2002O}. Typical searches for gravitational wave burst signals reconstruct signal waveforms using a combination of wavelets \cite{2015CQGra..32m5012C, 2008CQGra..25k4029K, PhysRevD.95.104046}. However, as CCSN signals contain multiple complex features in the time-frequency domain, these techniques often struggle to reconstruct the entire signal \cite{McIver:2015pms}. Previous searches for CCSN gravitational wave signals did not find any significant detection candidates \cite{2016PhRvD..94j2001A}.

In recent years, techniques such as principal component analysis (PCA) have been applied to CCSN gravitational waveforms to synthesize approximate models from the common features of the simulated waveforms, which can be used for model selection and waveform reconstruction. PCA is a technique for identifying common patterns in high-dimensional data and expressing the data to accentuate the similarities and differences in these patterns. This reduces, often significantly, the number of dimensions in a data set without much information loss. PCA was applied first to CCSN waveforms by Heng \cite{heng:09, 2009PhRvD..80j2004R}, who analysed the two-dimensional (2D) simulations of Dimmelmeier et al. \cite{dimmelmeier:08}. This work was extended by Logue et al. \cite{logue:12} to produce waveform models for different types of CCSN explosion mechanisms. This method is effective in real gravitational wave detector noise and reduces false alarms triggered by transient detector noise artefacts \cite{powell:16b, 2017PhRvD..96l3013P}.  

PCA has some limitations in its ability to reconstruct gravitational waveforms and hence elucidate the source physics. For example, only those segments  of the waveforms that are shorter than the longest waveform in the training set  can be reconstructed. Most simulations of CCSN are terminated early due to computational cost, resulting in waveforms which are shorter than what ground based detectors would actually measure. Furthermore, PCA does not recognize  when two signals have the same underlying functional form on a different time scale; it treats them as two different patterns instead of time-stretched versions of a single feature. 

To address the above limitations, we apply a technique known as dynamic time warping (DTW). DTW stretches localized segments of a time series in order to provide a better match to the pattern of interest. DTW and its modifications are widely used in  applications such as speech recognition~\cite{myers1980performance, myers1981comparative, muller2007dynamic}, human activity classification~\cite{sempena2011human,ten2007multi}, signature recognition~\cite{martens1996line} and shape matching~\cite{bartolini2005warp}. In this paper, we present a proof-of-principle study to demonstrate how this technique allows for better reconstruction of a CCSN gravitational waveform when it is applied before PCA. We show that DTW reduces the variance of a set of supernova waveforms, leading to faster and more robust analysis of CCSN signals in real gravitational wave data.   

In Section \ref{sec:supernova}, we provide a brief overview of gravitational wave signals from CCSNe and the waveforms that are used in this study. In Section \ref{sec:pca}, we describe the PCA algorithm. The DTW technique is described in Section \ref{sec:dtw}. We show how DTW reduces the minimum sufficient number of principal components (PCs) in Section \ref{sec:num_pcs}. In Section \ref{sec:analysis}, we quantify how much improvement in waveform reconstruction is gained by applying DTW. We show the reconstruction results in Section \ref{sec:results}. A discussion of the results and future work is given in Section \ref{sec:discussion}. 

\section{Supernova Models}
\label{sec:supernova}

\begin{figure}[ht]
\includegraphics[width=\columnwidth]{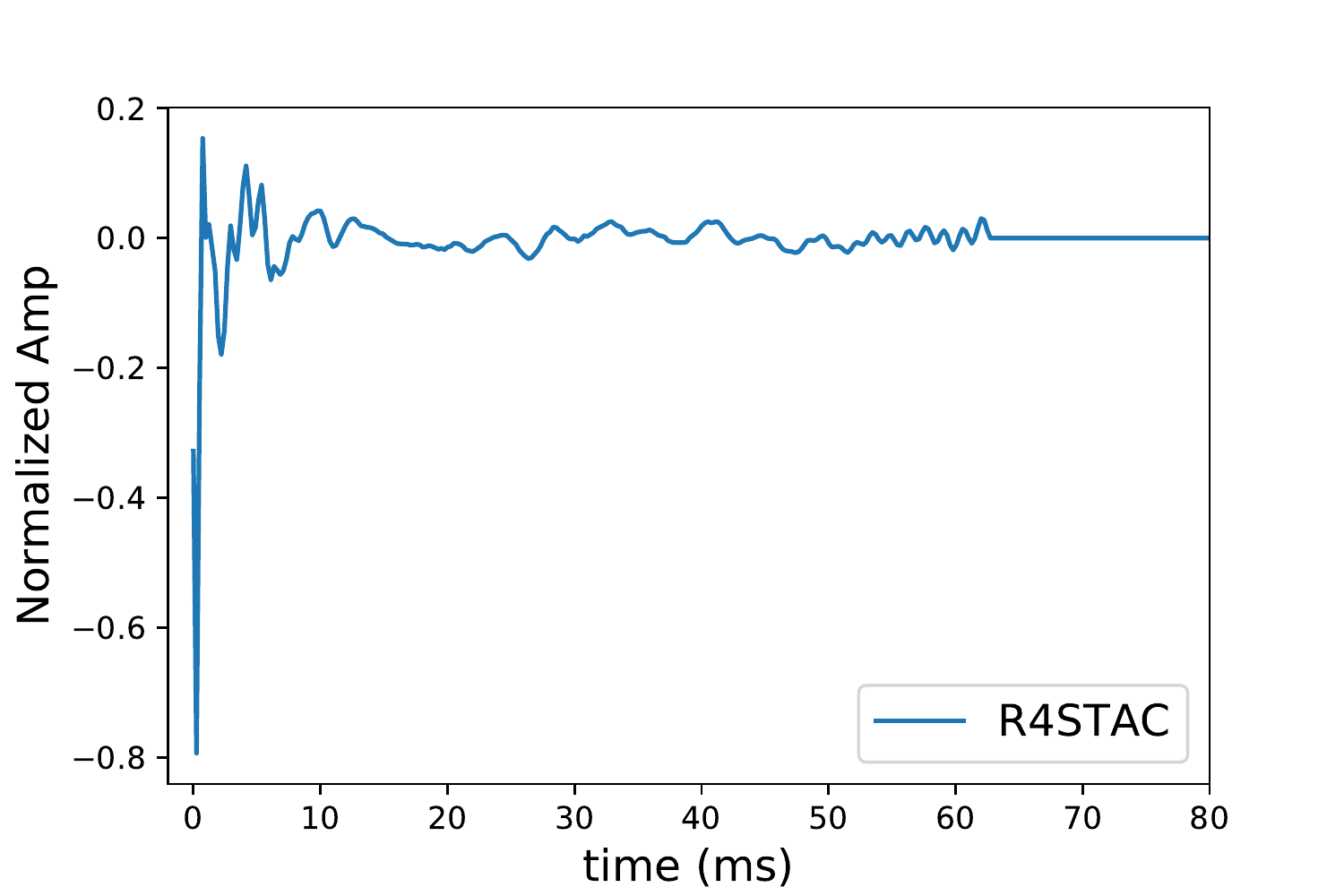}
\caption{An example waveform produced by Scheidegger et al. \cite{scheidegger2008gravitational}. A large spike occurs at core bounce due to the rapid rotation of the progenitor.}
\label{fig:waveforms}
\end{figure}

When a CCSN explodes, the shock stalls at a radius of $\sim150\,\mathrm{km}$, until an additional energy source revives it. The prevailing theory of shock revival is the neutrino-driven mechanism \cite{2012ARNPS..62..407J}. A number of three-dimensional (3D) CCSN gravitational wave neutrino-driven emission models have recently become available \cite{andresen:16, 2018arXiv181007638A, 2017arXiv170107325Y, kuroda:16, oconnor_18}. They display common features such as g-mode oscillations of the proto-neutron star (PNS) surface and the SASI \cite{0004-637X-584-2-971, 2006ApJ...642..401B, 2007ApJ...654.1006F}. However, the latest 3D neutrino-driven waveforms still differ widely in their predicted gravitational wave amplitudes and the frequencies at which the emission peaks.  

Another potential mechanism for shock revival is known as the magneto-rotational mechanism \cite{scheidegger2008gravitational}. The explosion in this case is powered by rapid rotation of the PNS, inherited from the progenitor, which is expected to be a relatively rare scenario \cite{2012Natur.481...55B, 2012A&A...548A..10M}.  In this study, we use the 25 magneto-rotational CCSN waveforms generated by Scheidegger {\it et al.} in~\cite{scheidegger2008gravitational}, as they are more plentiful than those from 3D neutrino-driven simulations. The waveforms include two orthogonal polarizations, $h_+$ and $h_\times$. 

A list of the parameters of the 25 waveforms is given in Table~\ref{tab:1}. The waveforms are simulated with a $15\,M_{\odot}$ zero age main sequence (ZAMS) progenitor using two different equations of state and various rotation rates. They typically last $\sim 200\,{\rm ms}$ with strain $\sim10^{-21}$ at a distance of 10\,kpc. A typical example waveform is displayed in Figure~\ref{fig:waveforms}. Most of the waveforms feature a large spike at core bounce followed by weaker emission. Although only one type of CCSN waveform is used in this study, we expect similar results for other waveforms.

\begin{table*}[ht]
\begin{centering}
\begin{tabular}{| c | c | c | c | c | c | c | c |}
  \hline
model & $t_\text{stop}$(ms) & $E_{GW}(M_{\odot}c^2)$& freq (Hz) & model & $t_\text{stop}$(ms) & $E_{GW}(M_{\odot}c^2)$& freq (Hz) \\ \hline
$R0E1CA$   & 130.8 & $2.15\times 10^{-11}$ & -   & $R3E3AC$   & 57.5  & $6.44\times 10^{-8}$ & 891 \\ 
$R0E3CA$   & 103.9 & $5.72\times 10^{-11}$ & -   & $R3STAC$   & 50.2  & $1.05\times 10^{-7}$  & 854 \\ 
$R0STCA*$   & 70.3  & $1.35\times 10^{-11}$ & -    & $R3E1CA$   & 69.4  & $8.05\times 10^{-8}$  & 897 \\ 
$R1E1CA$   & 112.8 & $1.36\times 10^{-10}$ & -   & $R3E1DB$   & 62.7  & $7.76\times 10^{-8}$  & 886 \\ 
$R1E3CA**$   & 130.4 & $1.43\times 10^{-10}$ & -     & $R3E1AC_L$ & 196.7 & $2.14\times 10^{-7}$  & 909 \\ 
$R1E1DB$   & 112.8 & $1.24\times 10^{-10}$ & -   & $R4E1AC$   & 98.7  & $7.74\times 10^{-8}$  & 385 \\ 
$R1STCA$   & 45.8  & $2.01\times 10^{-11}$ & -   & $R4STAC$   & 67.2  & $1.91\times 10^{-7}$  & 396 \\ 
$R1E1CA_L$ & 92.9  & $1.04\times 10^{-10}$ & -   & $R4E1EC*$   & 100.8 & $7.51\times 10^{-8}$  & 866 \\ 
$R2E1AC$   & 127.3 & $5.52\times 10^{-9}$  & 841 & $R4E1FC$   & 80.1  & $7.29\times 10^{-8}$  & 859 \\ 
$R2E3AC$   & 106.4 & $5.31\times 10^{-9}$  & 803 & $R4E1FC_L$ & 97.6  & $3.42\times 10^{-7}$  & 485 \\ 
$R2STAC$   & 64.0  & $7.62\times 10^{-9}$  & 680 & $R4E1CF$   & 19.1  & $6.50\times 10^{-8}$  & 370 \\ 
$R3E1AC*$   & 62.5  & $5.58\times 10^{-8}$  & 880 & $R5E1AC$   & 93.2  & $1.20\times 10^{-8}$  & 317 \\ 
$R3E2AC$   & 45.5  & $6.53\times 10^{-8}$  & 882 & & & \\ \hline
\end{tabular}
\caption[]{Summary of the properties of the waveforms used in this study reproduced from \cite{scheidegger:10b}. $t_\text{stop}$ is the time after core bounce when the simulation ends. $E_{GW}$ is the energy emitted in gravitational waves. The frequency refers to the peak of the strain spectrum at core bounce. The unstarred, **, and * entries refer to the training set, DTW template, and testing set respectively.}
\label{tab:1}
\end{centering}
\end{table*}

\section{Principal component analysis }
\label{sec:pca}

PCA represents a class of discrete time functions as a linear combination of orthogonal basis functions, known as PCs \cite{jolliffe2011principal}.  PCA removes the correlated variables, leaving behind a low-dimensional structure. The higher dimensions mostly contain noise. Conceptually, PCA is a way to find  a new coordinate system that reveals underlying linear relationships in the data.  Before applying PCA, the waveforms are stacked as rows of the data matrix $\bf{X}$ (zero padded if necessary), so that the dimension of the data matrix is $n\times T$, where $n$ is the number of waveforms, and $T$ is the waveform duration measured in time samples. The covariance is 
\begin{equation}
\frac{1}{n-1}(\bf{X} - \bar{\bf{X}})(\bf{X} - \bar{\bf{X}})^T = \bf{A}\bf{D}\bf{A}^T,
\end{equation}
where $\bar{\bf{X}}$ is the mean of each column of $\bf{X}$,  $\bf{D}$ is a diagonal matrix of eigenvalues sorted in descending order, and $\bf{A}$ is a matrix of eigenvectors. The PCs are  computed as 
\begin{equation}
\bf{P}= \bf{X}\bf{A}^T.
\end{equation}
The leading PCs contain most of the information in the data. 

PCA offers the following advantages: it accomodates data compression, highlights similarities and differences in patterns in the data, and removes any redundant features \cite{jolliffe2011principal}. It fails when the mean and covariance are not enough to characterize the data, for example when the quasirandom signal (not noise) components do not follow a normal distribution. There is no unique recipe for deciding how many PCs to keep; in practice, rules of thumb are applied~\cite{peres2005many}.

The truncated transformation $\bf{P}_L$ composed of the $L$ leading PCs compresses the data and guards against overfitting. A waveform $x$ is reconstructed into the basis of $\bf{P}_L$ as follows:
\begin{equation}
\label{eq:rec}
x_L={\bf P_L}{\bf P_L}^T x.
\end{equation}

The explained variance of the data is defined as,
\begin{equation}
v(k) = \frac{1}{\Sigma} \sum_{i=1}^{k} {D}_{i}, 
\label{eqn:var}
\end{equation}
with
\begin{equation}
\Sigma = \sum_{i=1}^{n} {D}_{i},
\end{equation}
where ${D}_{i}$ are the diagonal elements of matrix ${\bf D}$, $k$ is the number of PCs of interest, and $n$ is the total number of PCs. The explained variance measures the dispersion of the data set as a function of its dimensionality. 

In this study, we apply PCA to both polarizations of a subset of the waveforms described in Section \ref{sec:supernova} after normalizing and applying a pre-whitening filter \cite{karhunen1998nonlinear}. We select a subset of $22$ waveforms to train the PCA. The remaining waveforms are used for testing (* entries in Table~\ref{tab:1}). One training waveform (** entry in Table~\ref{tab:1}) also serves as a DTW template. In this work the training waveforms, DTW template, and test waveforms are selected arbitrarily. 

\begin{figure*}[ht]
\includegraphics[height=10cm]{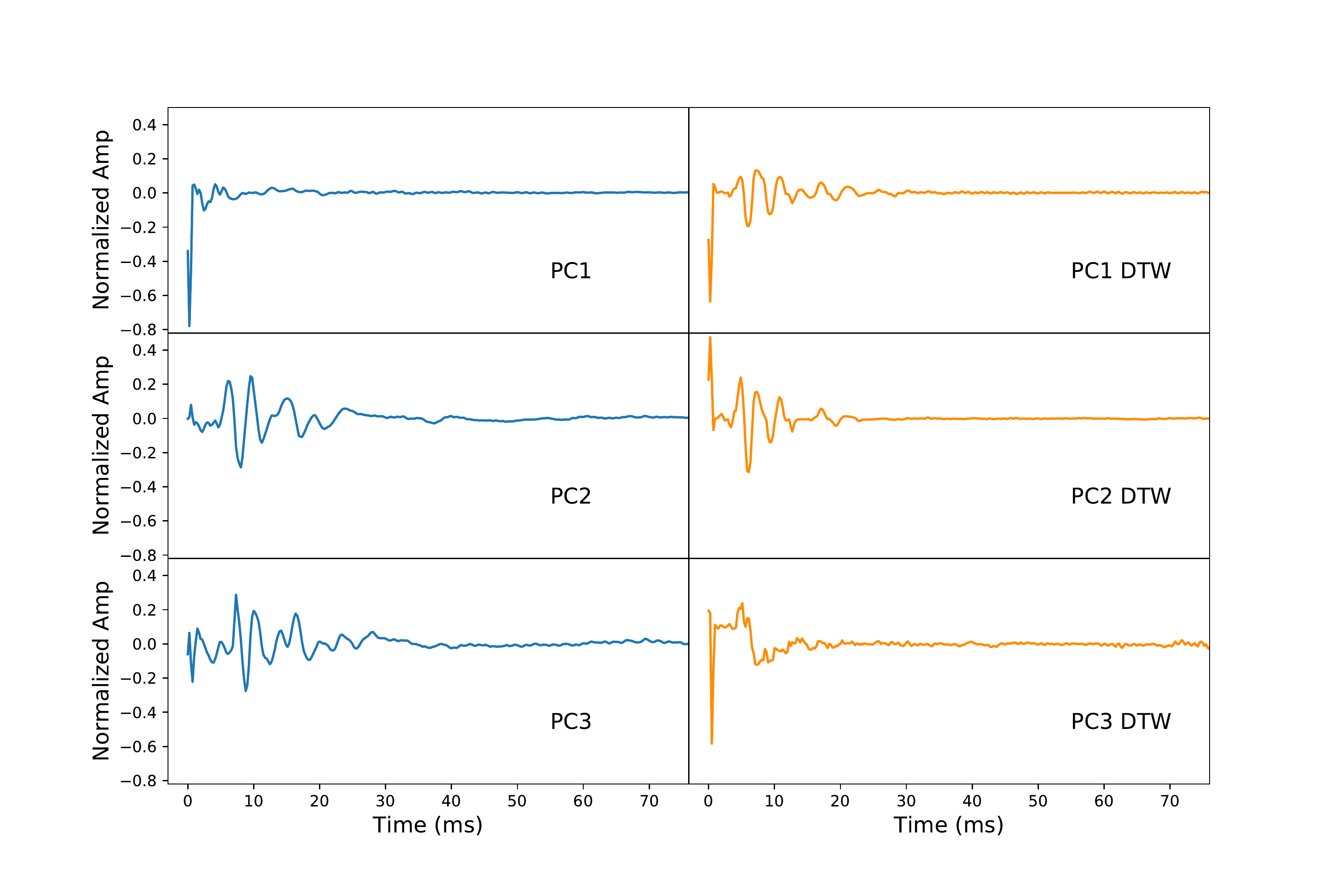}
\caption{The three leading PCs produced with the original waveforms (left) and time warped waveforms (right). The spike at core bounce is more pronounced in the time warped PCs.
\label{fig:pcs} }
\end{figure*}

The first three PCs, before the application of DTW, are shown in the left column of Figure \ref{fig:pcs}. The main feature in the first PC is the spike at core bounce.   


PCA was applied to the waveforms from Scheidegger et al. \cite{scheidegger:10b} in Ref.\ \cite{2017PhRvD..96l3013P} and other types of waveforms in Ref.\ \cite{powell:16b, logue:12}. However, a real CCSN waveform may not match exactly what a simulation predicts. Many simulations are stopped early due to computational cost, and PCA models cannot reconstruct any waveform longer than the simulations. We aim to reduce the number of PCs as far as possible to accelerate the analysis when it is combined with nested sampling as in previous studies \cite{2017PhRvD..96l3013P, powell:16b, logue:12}. These issues and others can be addressed with the aid of DTW. 

\section{Dynamic time warping}
\label{sec:dtw}


A good way to align two signals is with a dynamic programming algorithm known as dynamic time warping (DTW). The earliest reference to the method is found in ~\cite{sakoe1978dynamic}. 

For two discretely sampled time domain signals ${\bf f}=(f_1,\ldots,f_N)$ and ${\bf g}=(g_1,\ldots,g_M)$, the DTW algorithm finds two discrete warping functions $w(t)=(w_1,w_2,\ldots,w_K)$ and $u(t)=(u_1,u_2,\ldots,u_K)$ such that the distance
$\sum\limits_{i=1}^K|f_{w_i}-g_{u_i}|$ is minimized, where the brackets denote Euclidean distance (or any other metric of choice) between the two signals. The warping functions $w(t)$ and $u(t)$ each correspond to an ordered sequence of instants, where the signals $f(t)$ and $g(t)$ are sampled respectively. 



To align multiple signals with a given template $\bf{p}$, a warping $\mu_n(t)=w^{(n)-1}[u^{(n)}(t)]$ is computed,  where $w^{(n)}$ and $u^{(n)}$ are the optimal warping functions between $\bf{p}$ and  the $n$-th waveform ${\bf f}_n$ (not the template) in a database. Here the superscript $-1$ denotes the inverse function, not the reciprocal.  The warped signals ${\bf g}_n ={\bf f}_n [\mu_n(t)]$ are aligned with the template $\bf{p}$ and themselves.  The inverse transformation $\mu_n^{-1}(t)=u^{(n)-1}[w^{(n)}(t)]$  unwarps ${\bf g}_n$, transforming it  back to ${\bf f}_n$. Both warp and unwarp transformations result in some loss of time-series information. However, in our case the loss due to DTW is typically smaller than the information lost due to the detector noise. 

Briefly, DTW works as follows.
For two discretely sampled signals ${\bf f}=(f_1,\ldots, f_N)$ and ${\bf g}=(g_1,\ldots g_M)$ a distance matrix $D$ is computed
\begin{equation}
D=\begin{pmatrix} |f_1-g_1| & |f_1-g_2| & \ldots |f_1-g_M|\\
|f_2-g_1| & |f_2-g_2| & \ldots |f_2-g_M|\\
\cdots\\
|f_N-g_1| & |f_N-g_2| & \ldots |f_N-g_M|\\
\end{pmatrix}. 
\end{equation}
Dynamic programming is then used to find a continuous path from the top left corner $D_{11}$ to the bottom right corner $D_{NM}$ of the matrix that minimizes the sum of the elements along the path. To ensure that the mapping is monotonic and nondecreasing, only three  movements along a path are  allowed:  one element  down, one right, or one diagonal right. 
%
An important point physically is that DTW stretches the time coordinate locally, not globally. In terms of the template, for example, the DTW effectively finds the optimal local scaling factor $\mu(t)$, such that $f_n(t)$ is stretched timewise into $g_n(t)=f_n[\mu(t)]$.


In our work, the template waveform is arbitrarily chosen to be model R1E3CA. A preliminary investigation shows that most of the signals in Table \ref{tab:1} (except R3E2AC and R4E1CF, which are idiosyncratic) are suitable templates.

\begin{figure*}[ht]
\includegraphics[width=0.67\textwidth,height=6.5cm]{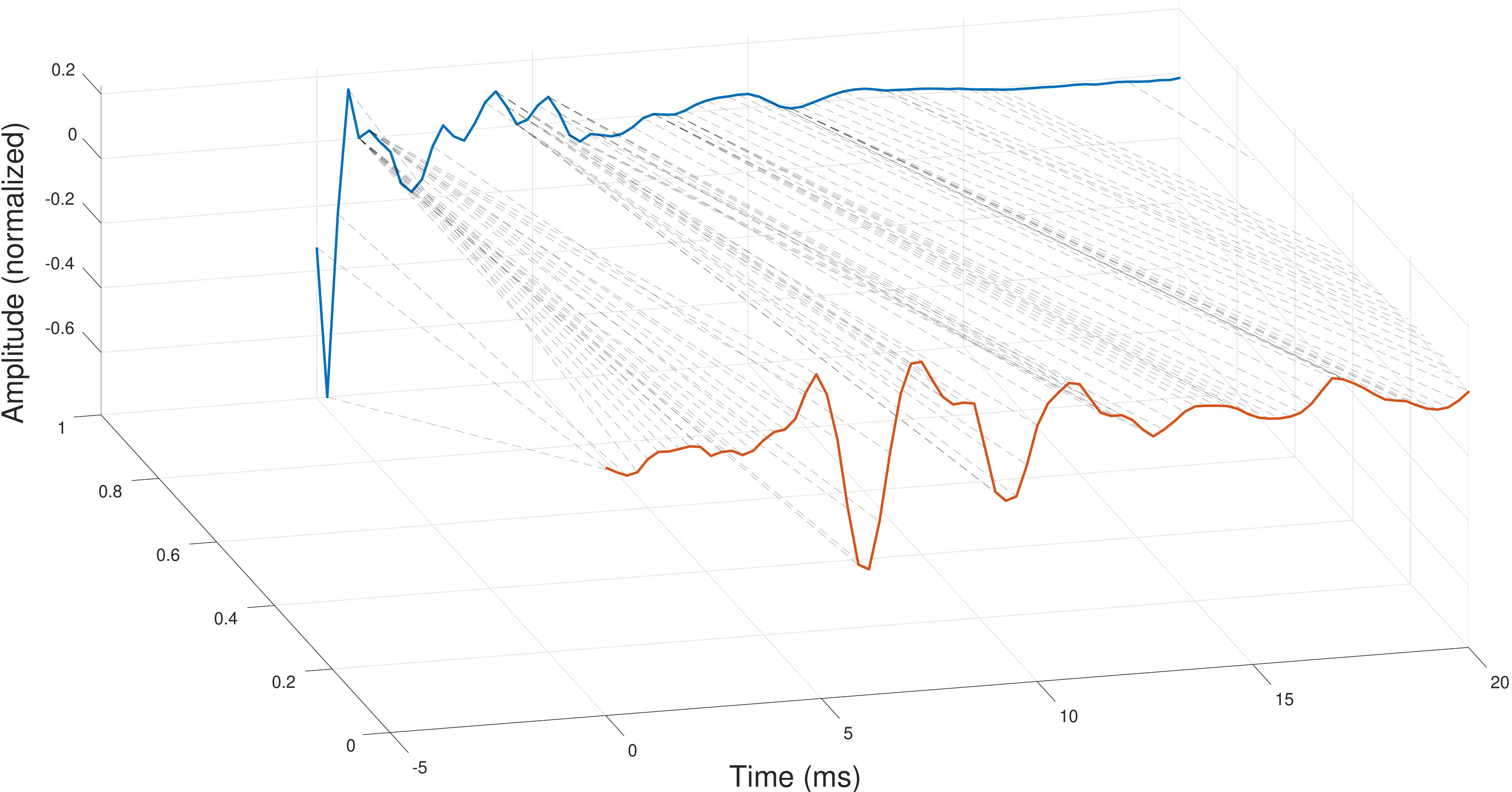}
\caption{Illustrative example of the application of DTW to one waveform. The blue waveform is model R3E1AC. The red waveform is the template R2E1AC. The black lines map the warping of R3E1AC to R2E1AC. \label{fig:w2}}
\end{figure*}

To improve the performance of PCA CCSN analysis, we first apply DTW to the pre-whitened waveforms. Some information is lost after DTW, but the variance of the waveforms has now been reduced. A more detailed example of how one waveform is warped with respect to the template waveform is shown in Figure \ref{fig:w2}. After DTW, we apply PCA to the warped signals. The original waveforms are aligned at core bounce before applying PCA, but when DTW is applied the alignment is made automatically. The first three PCs for the DTW signals are shown in Figure \ref{fig:pcs}. The spike at core bounce is more pronounced in the warped PCs than in the original PCs.  

\section{Number of Principal Components}
\label{sec:num_pcs}

DTW reduces the differences between original waveforms, so fewer PCs should be needed to represent the data accurately. To determine if this is the case, we look at the explained variance of the data, as given by Equation \ref{eqn:var}. The explained variance quantifies how much of the total variance is represented by each PC. The larger the explained variance, the better the PCs match the data. 

\begin{figure}[ht]
\includegraphics[width=\columnwidth]{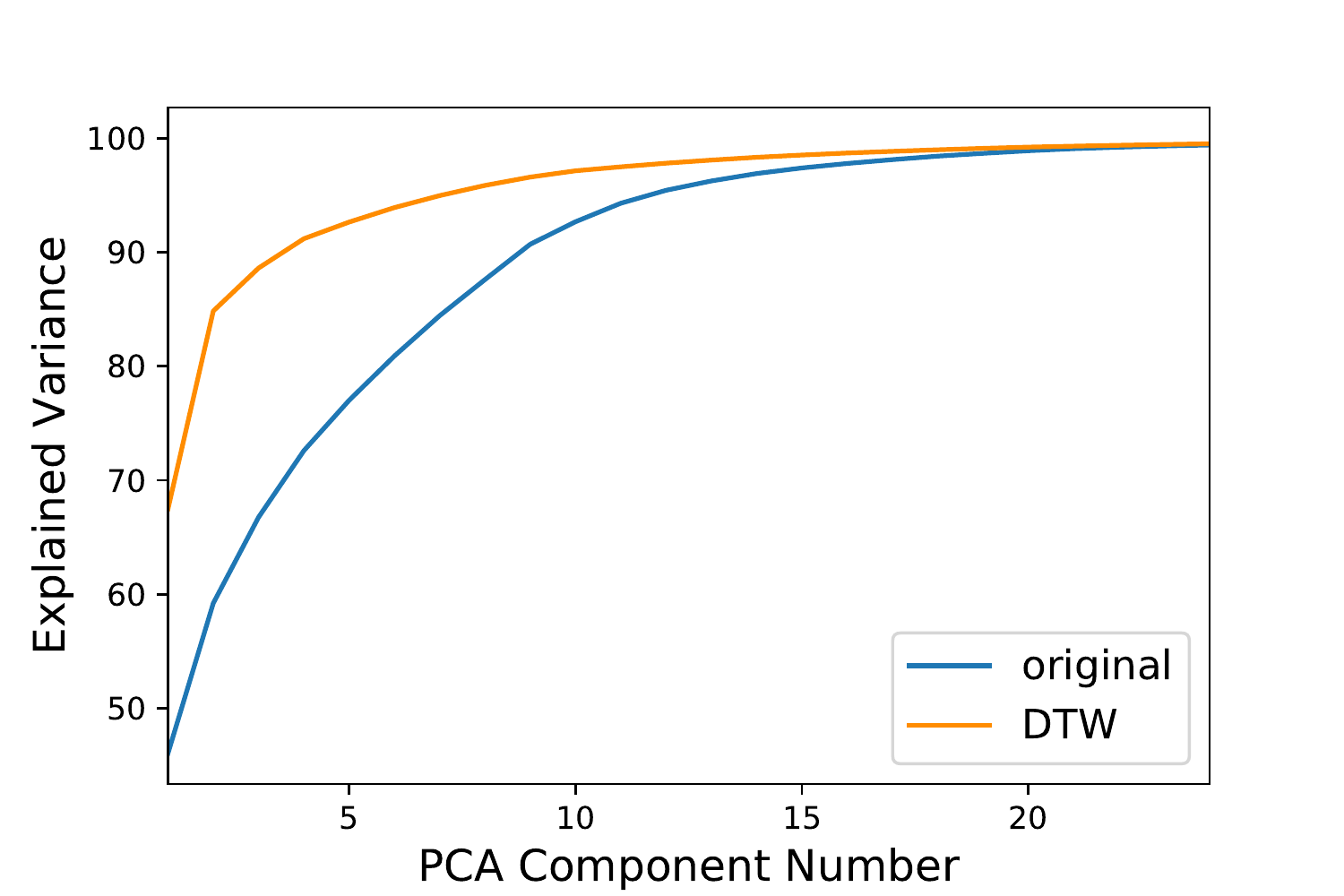}
\caption{Cumulative variance as a function of the number of PCs. Nine of the original PCs capture 90\% of the variance. When DTW is applied before computing the PCs, four PCs capture 90\% of the variance.
\label{fig:var}}
\end{figure}

Figure~\ref{fig:var} shows the explained variance for the waveforms in the training set versus the number of PCs for the original and time-warped waveforms. The number of PCs needed to represent 90\% of the variance of the data is nine for the original PCs and four for the DTW PCs, a clear reduction. 

Reducing the number of PCs is important for the Bayesian model selection techniques applied to CCSN waveforms in \cite{2017PhRvD..96l3013P}, because the analysis is faster when fewer PCs are used, and Bayesian model selection punishes excessively complicated models. However, we need to check whether or not DTW compromises signal reconstruction. Reducing the number of PCs is only effective, if information about the reconstructed signal is preserved. To test the effects of DTW on reconstruction, the leading PCs must be combined linearly to produce a model for the signal in the noise as described in Equation (\ref{eq:rec}). To do this below, we use the first nine original PCs and four DTW PCs. 

\section{Reconstruction Method}
\label{sec:analysis}


In this section, we describe the stages in the test process applied to determine how well DTW improves waveform reconstruction. As this paper is a proof-of-principle study, we consider only a single Advanced LIGO detector. We design the pre-whitenning filter for the Gaussian noise created using the zero detuning, high power, noise power spectral density (PSD) for Advanced LIGO at design sensitivity \cite{2018LRR....21....3A}, into which the three test signals described in Section \ref{sec:supernova} are injected. We compare reconstructions using the original and DTW PCs.   

First, we form the test signals according to
\begin{equation}
\label{eq:received}
h(t)=F_+h_+(t)+F_\times h_\times (t) + \nu(t),
\end{equation}
where $h_+(t)$ and $h_\times(t)$ are two independent polarization amplitudes, 
$F_+$ and $F_\times$ are the antenna patterns, and
$\nu(t)$ is zero mean noise with covariance $\Sigma(t,s)=E[\nu(t)\nu(s)]$.
Owing to the short duration of the waveforms, it is appropriate to assume that both antenna pattern functions are real constants, and that the noise is Gaussian and stationary. We take $F_+=F_\times=0.5$ by way of illustration. The distance of the signals is varied to achieve a signal-to-noise ratio (SNR) between 0 and 50. The matched filter SNR is given by
\begin{equation}
{\rm SNR}^{2} = 4 \int^{\infty}_{0} df\, \frac{|h(f)|^{2}}{S(f)}~,
\end{equation}
where $S(f)$ is the one-sided PSD. 

For the reconstructions produced using the original PCs, the pre-whitened noisy data is projected onto the PC basis comprising the first nine PCs as described in Section \ref{sec:num_pcs}. The projection generates a set of PC coefficients, which let us reconstruct the signal using Equation (\ref{eq:rec}).

For the DTW case, the next stage is to find the optimal warping between the noisy whitened data and the (arbitrarily selected) $+$ polarization of the template waveform. The resulting warped signal is projected onto the PC basis, the inverse unwarp transform is applied to the PCs, and the waveform is reconstructed and compared with the original waveform. 

We evaluate the accuracy of the reconstruction in terms of the match score, 
\begin{equation}
M = \frac{({\tilde{h}}|{\bar{h}})}{\sqrt{ ({\tilde{h}}|{\tilde{h}})({\bar{h}}|{ \bar{h}}) }}
\end{equation}
where ${\tilde{h}}$ is the pre-whitened received signal,  ${ \bar{h}}$ is the reconstructed waveform, and $(a|b)$ denotes the noise-weighted
 inner product.  
The match parameter is calculated using all test signals and $10^3$ realizations of the noise. 


\section{Reconstruction Results}
\label{sec:results}

\begin{figure*}[ht]
\includegraphics[width=\textwidth]{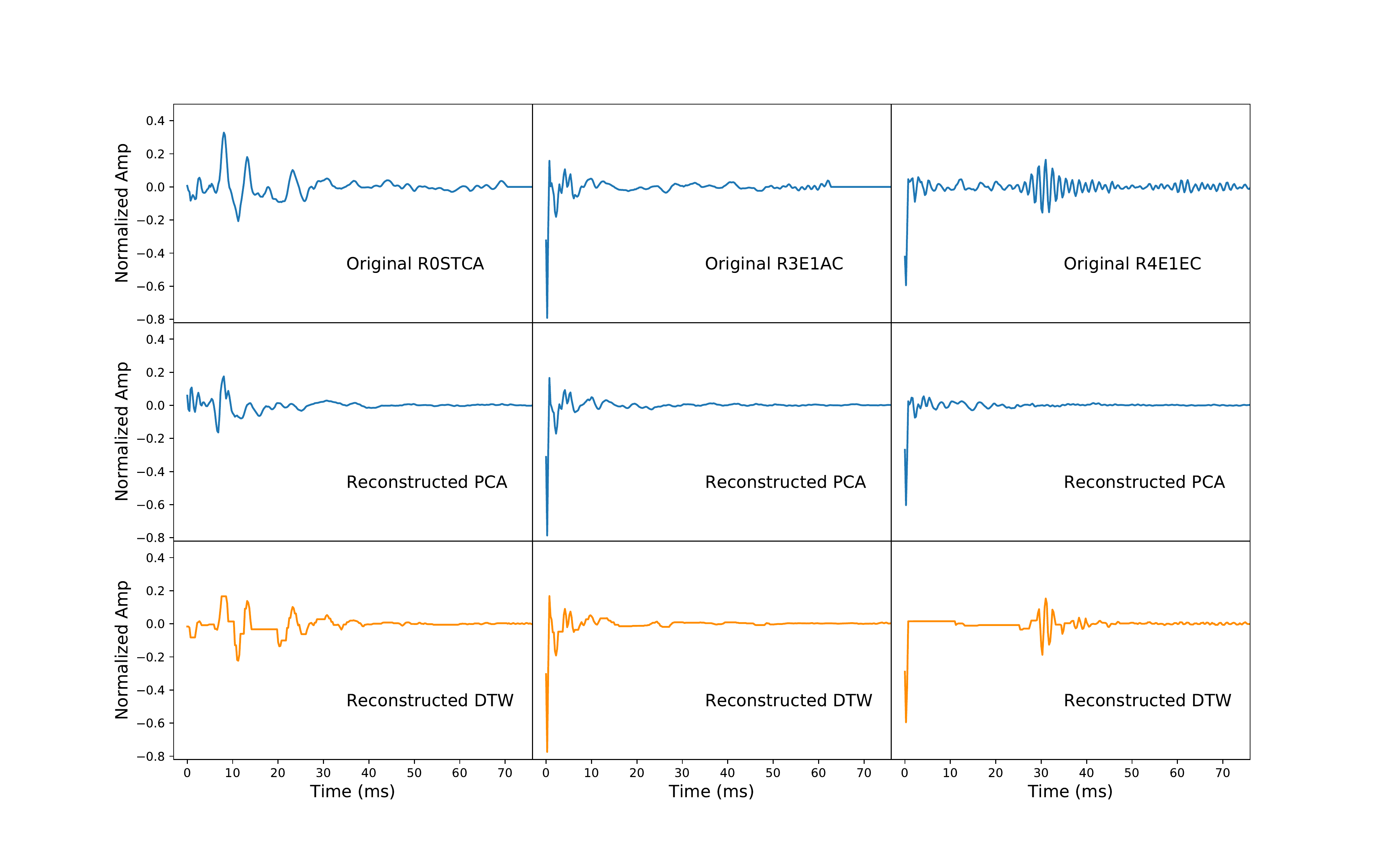}
\caption{Example reconstructions for the three test waveforms used in this study. Top row: original waveforms. Middle row: reconstructions using the original PCs. Bottom row: reconstuctions after DTW is applied. DTW clearly captures low amplitude features in the tails of the waveforms which are missed by the original PCs.
\label{fig:5a}}
\end{figure*}

\begin{figure*}[ht]
\includegraphics[width=0.45\textwidth,height=5cm]{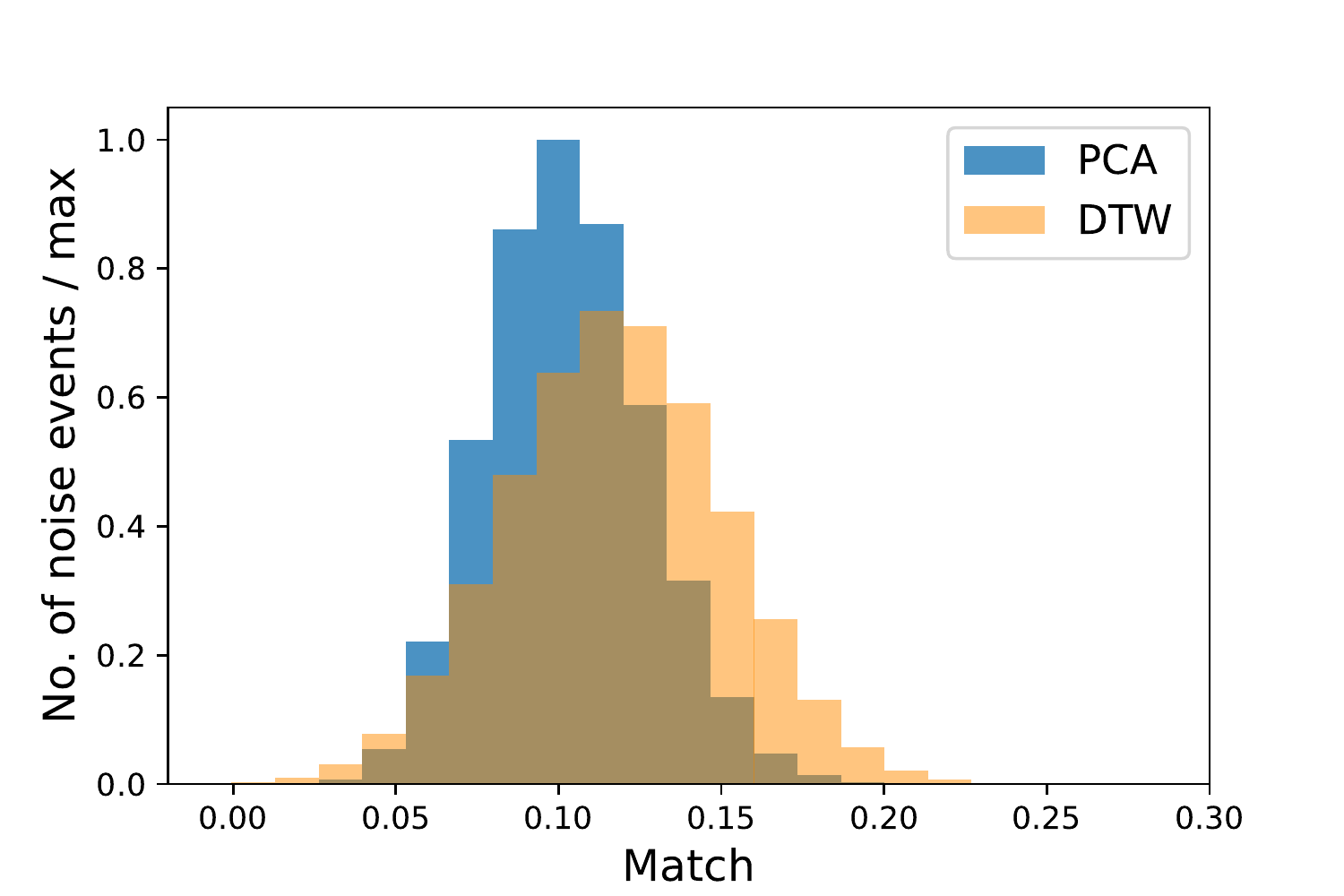}
\includegraphics[width=0.45\textwidth,height=5cm]{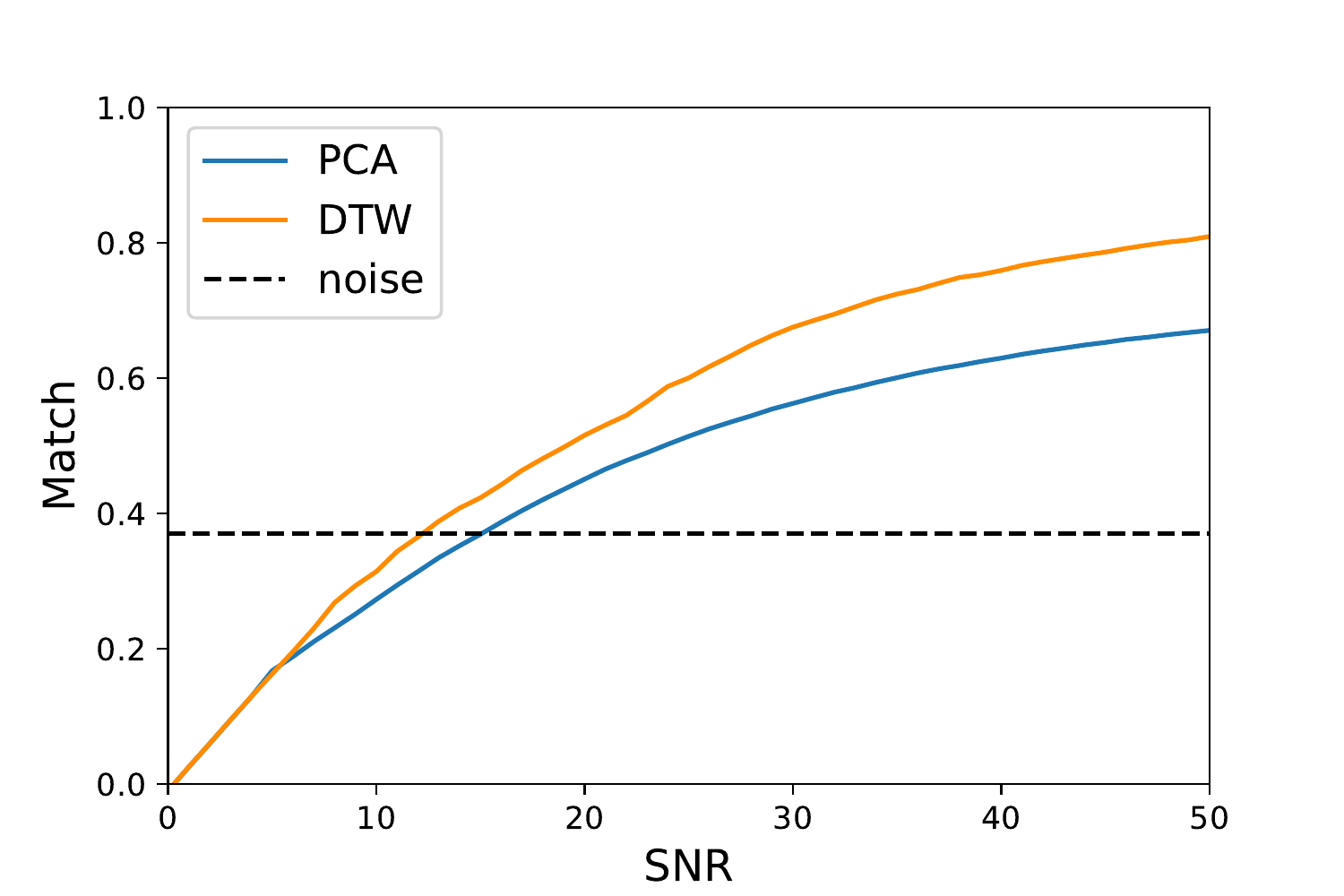}
\caption{(Left) Histogram of the match score $M$ for noise-only realizations with the original (blue) and DTW (orange) PCs. (Right) Average match score for the three test signals versus SNR. Applying DTW improves the reconstructions at all SNR values and reduces the minimum SNR. 
\label{fig:5}}
\end{figure*}


One example reconstructed waveform for each of the three test waveforms is shown in Figure~\ref{fig:5a}. Both standard PCA and DTW PCA reconstruct the waveform accurately around core bounce, as this feature is heavily represented in the first few PCs for both methods. However, during the lower amplitude portions of the signals at later times, DTW method clearly does better at reconstructing subtle features.    

By way of calibration, we calculate the match scores for noise-only realizations without injected signals. It is important to check that we are not reconstructing things so well that we can now make noise look like signals. The results are shown in the left panel of Figure~\ref{fig:5}. Although $M$ does increase slightly, when DTW is applied, it is still significantly lower than what we would expect from a real gravitational wave signal. We calculate a threshold $\eta$ from the cumulative distribution function of the match score for noise-only realizations $P_\text{noise}$,  such that we have 
\begin{equation}
1-P_\text{noise}(\eta)  \le \text{FAP}.
\end{equation}  
%
For the false alarm probability $\text{FAP}=1.2\times 10^{-5}$, we find $\eta=0.37$. 
   
For $0 \leq {\rm SNR} \leq 50$, the average $M$ between the original waveform and the reconstruction of the noisy waveform is computed for all waveforms in the test set and $10^3$ noise realizations. This is displayed as a function of the SNR for the original PCA and DTW PCA methods in Figure~\ref{fig:5}. We find that CCSN waveform reconstructions rise above the noise threshold at ${\rm SNR}=11$ for the DTW PCA, as compared to ${\rm SNR} =16$ for the original PCA. 

Figure~\ref{fig:5} shows that $M$ for the DTW PCs exceeds $M$ for the original PCs across all the SNR values tested here. The improved reconstructions should lead to better model selection, if the method is incorporated into studies such as those in \cite{powell:16b, 2017PhRvD..96l3013P}. The maximum $M$ obtained is 0.67 for the original PCs and 0.80 for the DTW PCs. The maximum $M$ would likely increase, if more sophisticated noise reduction methods were combined with our analysis, such as those in \cite{2017PhRvD..96l3013P, 2015CQGra..32m5012C, 2008CQGra..25k4029K}.

\section{Discussion}
\label{sec:discussion}

In this work, we enhance the PCA method used in classification of CCSN waveforms by introducing DTW. We evaluate the proposed method on a library of simulated CCSN gravitational waveforms \cite{scheidegger:10b}. We show that DTW reduces the number of PCs needed to represent 90\% of the variability in the data from nine to just four. This accelerates the analysis, when DTW PCA is combined with techniques such as nested sampling. 

We simulate Gaussian noise characteristic of Advanced LIGO at design sensitivity and inject waveforms $10^3$ times, at different SNRs, to determine how well a signal can be reconstructed. We show that the minimum SNR needed to lift the match score above the noise threshold is reduced, and a higher maximum match score is achievable, when DTW is applied. 

In future work, the method can be extended to other models of CCSN waveforms, which in turn may help identify the explosion mechanism. The method can be incorporated into existing CCSN search algorithms, such as those described in \cite{powell:16b, 2017PhRvD..96l3013P}, as well as in other areas of gravitational wave data analysis, e.g.\ matched filter searches, to reduce the size of template banks. DTW may also prove useful for classifying noise transients \cite{2015CQGra..32u5012P, 2017CQGra..34c4002P}.
  
\begin{acknowledgments}
We thank Ik Siong Heng for useful discussions related to this work. 
The authors are supported 
by the Australian Research Council Centre of Excellence for Gravitational Wave Discovery 
(OzGrav), through project number CE170100004. 
\end{acknowledgments}

%
%

\begin{thebibliography}{46}
\expandafter\ifx\csname natexlab\endcsname\relax\def\natexlab#1{#1}\fi
\expandafter\ifx\csname bibnamefont\endcsname\relax
  \def\bibnamefont#1{#1}\fi
\expandafter\ifx\csname bibfnamefont\endcsname\relax
  \def\bibfnamefont#1{#1}\fi
\expandafter\ifx\csname citenamefont\endcsname\relax
  \def\citenamefont#1{#1}\fi
\expandafter\ifx\csname url\endcsname\relax
  \def\url#1{\texttt{#1}}\fi
\expandafter\ifx\csname urlprefix\endcsname\relax\def\urlprefix{URL }\fi
\providecommand{\bibinfo}[2]{#2}
\providecommand{\eprint}[2][]{\url{#2}}

\bibitem[{\citenamefont{{The LIGO Scientific Collaboration}
  et~al.}(2015)\citenamefont{{The LIGO Scientific Collaboration}, {Aasi},
  {Abbott}, {Abbott}, and et~al.}}]{aLIGO}
\bibinfo{author}{\bibnamefont{{The LIGO Scientific Collaboration}}},
  \bibinfo{author}{\bibfnamefont{J.}~\bibnamefont{{Aasi}}},
  \bibinfo{author}{\bibfnamefont{B.~P.} \bibnamefont{{Abbott}}},
  \bibinfo{author}{\bibfnamefont{R.}~\bibnamefont{{Abbott}}}, \bibnamefont{and}
  \bibinfo{author}{\bibnamefont{et~al.}}, \bibinfo{journal}{\cqg}
  \textbf{\bibinfo{volume}{32}}, \bibinfo{eid}{074001} (\bibinfo{year}{2015}),
  \eprint{1411.4547}.

\bibitem[{\citenamefont{{Acernese} and et~al.}(2015)}]{AdVirgo}
\bibinfo{author}{\bibfnamefont{F.}~\bibnamefont{{Acernese}}} \bibnamefont{and}
  \bibinfo{author}{\bibnamefont{et~al.}}, \bibinfo{journal}{\cqg}
  \textbf{\bibinfo{volume}{32}}, \bibinfo{eid}{024001} (\bibinfo{year}{2015}),
  \eprint{1408.3978}.

\bibitem[{\citenamefont{{Gossan} et~al.}(2016)\citenamefont{{Gossan}, {Sutton},
  {Stuver}, {Zanolin}, {Gill}, and {Ott}}}]{2016PhRvD..93d2002G}
\bibinfo{author}{\bibfnamefont{S.~E.} \bibnamefont{{Gossan}}},
  \bibinfo{author}{\bibfnamefont{P.}~\bibnamefont{{Sutton}}},
  \bibinfo{author}{\bibfnamefont{A.}~\bibnamefont{{Stuver}}},
  \bibinfo{author}{\bibfnamefont{M.}~\bibnamefont{{Zanolin}}},
  \bibinfo{author}{\bibfnamefont{K.}~\bibnamefont{{Gill}}}, \bibnamefont{and}
  \bibinfo{author}{\bibfnamefont{C.~D.} \bibnamefont{{Ott}}},
  \bibinfo{journal}{\prd} \textbf{\bibinfo{volume}{93}}, \bibinfo{eid}{042002}
  (\bibinfo{year}{2016}), \eprint{1511.02836}.

\bibitem[{\citenamefont{{Abbott} et~al.}(2016)\citenamefont{{Abbott}, {Abbott},
  {Abbott}, {Abernathy}, {Acernese}, {Ackley}, {Adams}, {Adams}, {Addesso},
  {Adhikari} et~al.}}]{2016PhRvD..94j2001A}
\bibinfo{author}{\bibfnamefont{B.~P.} \bibnamefont{{Abbott}}},
  \bibinfo{author}{\bibfnamefont{R.}~\bibnamefont{{Abbott}}},
  \bibinfo{author}{\bibfnamefont{T.~D.} \bibnamefont{{Abbott}}},
  \bibinfo{author}{\bibfnamefont{M.~R.} \bibnamefont{{Abernathy}}},
  \bibinfo{author}{\bibfnamefont{F.}~\bibnamefont{{Acernese}}},
  \bibinfo{author}{\bibfnamefont{K.}~\bibnamefont{{Ackley}}},
  \bibinfo{author}{\bibfnamefont{C.}~\bibnamefont{{Adams}}},
  \bibinfo{author}{\bibfnamefont{T.}~\bibnamefont{{Adams}}},
  \bibinfo{author}{\bibfnamefont{P.}~\bibnamefont{{Addesso}}},
  \bibinfo{author}{\bibfnamefont{R.~X.} \bibnamefont{{Adhikari}}},
  \bibnamefont{et~al.}, \bibinfo{journal}{\prd} \textbf{\bibinfo{volume}{94}},
  \bibinfo{eid}{102001} (\bibinfo{year}{2016}), \eprint{1605.01785}.

\bibitem[{\citenamefont{{M{\"u}ller}}(2017)}]{2017arXiv170304633M}
\bibinfo{author}{\bibfnamefont{B.}~\bibnamefont{{M{\"u}ller}}},
  \bibinfo{journal}{ArXiv e-prints}  (\bibinfo{year}{2017}),
  \eprint{1703.04633}.

\bibitem[{\citenamefont{{Blondin} et~al.}(2003)\citenamefont{{Blondin},
  {Mezzacappa}, and {DeMarino}}}]{2003ApJ...584..971B}
\bibinfo{author}{\bibfnamefont{J.~M.} \bibnamefont{{Blondin}}},
  \bibinfo{author}{\bibfnamefont{A.}~\bibnamefont{{Mezzacappa}}},
  \bibnamefont{and}
  \bibinfo{author}{\bibfnamefont{C.}~\bibnamefont{{DeMarino}}},
  \bibinfo{journal}{\apj} \textbf{\bibinfo{volume}{584}}, \bibinfo{pages}{971}
  (\bibinfo{year}{2003}), \eprint{astro-ph/0210634}.

\bibitem[{\citenamefont{{Andresen} et~al.}(2017)\citenamefont{{Andresen},
  {M{\"u}ller}, {M{\"u}ller}, and {Janka}}}]{andresen:16}
\bibinfo{author}{\bibfnamefont{H.}~\bibnamefont{{Andresen}}},
  \bibinfo{author}{\bibfnamefont{B.}~\bibnamefont{{M{\"u}ller}}},
  \bibinfo{author}{\bibfnamefont{E.}~\bibnamefont{{M{\"u}ller}}},
  \bibnamefont{and} \bibinfo{author}{\bibfnamefont{H.-T.}
  \bibnamefont{{Janka}}}, \bibinfo{journal}{\mnras}
  \textbf{\bibinfo{volume}{468}}, \bibinfo{pages}{2032} (\bibinfo{year}{2017}),
  \eprint{1607.05199}.

\bibitem[{\citenamefont{{M{\"u}ller} et~al.}(2012)\citenamefont{{M{\"u}ller},
  {Janka}, and {Wongwathanarat}}}]{mueller:e12}
\bibinfo{author}{\bibfnamefont{E.}~\bibnamefont{{M{\"u}ller}}},
  \bibinfo{author}{\bibfnamefont{H.-T.} \bibnamefont{{Janka}}},
  \bibnamefont{and}
  \bibinfo{author}{\bibfnamefont{A.}~\bibnamefont{{Wongwathanarat}}},
  \bibinfo{journal}{\aap} \textbf{\bibinfo{volume}{537}}, \bibinfo{eid}{A63}
  (\bibinfo{year}{2012}), \eprint{1106.6301}.

\bibitem[{\citenamefont{{Kuroda} et~al.}(2016)\citenamefont{{Kuroda}, {Kotake},
  and {Takiwaki}}}]{kuroda:16}
\bibinfo{author}{\bibfnamefont{T.}~\bibnamefont{{Kuroda}}},
  \bibinfo{author}{\bibfnamefont{K.}~\bibnamefont{{Kotake}}}, \bibnamefont{and}
  \bibinfo{author}{\bibfnamefont{T.}~\bibnamefont{{Takiwaki}}},
  \bibinfo{journal}{\apjl} \textbf{\bibinfo{volume}{829}}, \bibinfo{eid}{L14}
  (\bibinfo{year}{2016}), \eprint{1605.09215}.

\bibitem[{\citenamefont{{Morozova} et~al.}(2018)\citenamefont{{Morozova},
  {Radice}, {Burrows}, and {Vartanyan}}}]{morozova_18}
\bibinfo{author}{\bibfnamefont{V.}~\bibnamefont{{Morozova}}},
  \bibinfo{author}{\bibfnamefont{D.}~\bibnamefont{{Radice}}},
  \bibinfo{author}{\bibfnamefont{A.}~\bibnamefont{{Burrows}}},
  \bibnamefont{and}
  \bibinfo{author}{\bibfnamefont{D.}~\bibnamefont{{Vartanyan}}},
  \bibinfo{journal}{\apj} \textbf{\bibinfo{volume}{861}}, \bibinfo{eid}{10}
  (\bibinfo{year}{2018}), \eprint{1801.01914}.

\bibitem[{\citenamefont{{Yakunin} et~al.}(2017)\citenamefont{{Yakunin},
  {Mezzacappa}, {Marronetti}, {Lentz}, {Bruenn}, {Hix}, {Messer}, {Endeve},
  {Blondin}, and {Harris}}}]{2017arXiv170107325Y}
\bibinfo{author}{\bibfnamefont{K.~N.} \bibnamefont{{Yakunin}}},
  \bibinfo{author}{\bibfnamefont{A.}~\bibnamefont{{Mezzacappa}}},
  \bibinfo{author}{\bibfnamefont{P.}~\bibnamefont{{Marronetti}}},
  \bibinfo{author}{\bibfnamefont{E.~J.} \bibnamefont{{Lentz}}},
  \bibinfo{author}{\bibfnamefont{S.~W.} \bibnamefont{{Bruenn}}},
  \bibinfo{author}{\bibfnamefont{W.~R.} \bibnamefont{{Hix}}},
  \bibinfo{author}{\bibfnamefont{O.~E.~B.} \bibnamefont{{Messer}}},
  \bibinfo{author}{\bibfnamefont{E.}~\bibnamefont{{Endeve}}},
  \bibinfo{author}{\bibfnamefont{J.~M.} \bibnamefont{{Blondin}}},
  \bibnamefont{and} \bibinfo{author}{\bibfnamefont{J.~A.}
  \bibnamefont{{Harris}}}, \bibinfo{journal}{ArXiv e-prints}
  (\bibinfo{year}{2017}), \eprint{1701.07325}.

\bibitem[{\citenamefont{{Owen} and
  {Sathyaprakash}}(1999)}]{1999PhRvD..60b2002O}
\bibinfo{author}{\bibfnamefont{B.~J.} \bibnamefont{{Owen}}} \bibnamefont{and}
  \bibinfo{author}{\bibfnamefont{B.~S.} \bibnamefont{{Sathyaprakash}}},
  \bibinfo{journal}{\prd} \textbf{\bibinfo{volume}{60}}, \bibinfo{eid}{022002}
  (\bibinfo{year}{1999}), \eprint{gr-qc/9808076}.

\bibitem[{\citenamefont{{Cornish} and
  {Littenberg}}(2015)}]{2015CQGra..32m5012C}
\bibinfo{author}{\bibfnamefont{N.~J.} \bibnamefont{{Cornish}}}
  \bibnamefont{and} \bibinfo{author}{\bibfnamefont{T.~B.}
  \bibnamefont{{Littenberg}}}, \bibinfo{journal}{Classical and Quantum Gravity}
  \textbf{\bibinfo{volume}{32}}, \bibinfo{eid}{135012} (\bibinfo{year}{2015}),
  \eprint{1410.3835}.

\bibitem[{\citenamefont{{Klimenko} et~al.}(2008)\citenamefont{{Klimenko},
  {Yakushin}, {Mercer}, and {Mitselmakher}}}]{2008CQGra..25k4029K}
\bibinfo{author}{\bibfnamefont{S.}~\bibnamefont{{Klimenko}}},
  \bibinfo{author}{\bibfnamefont{I.}~\bibnamefont{{Yakushin}}},
  \bibinfo{author}{\bibfnamefont{A.}~\bibnamefont{{Mercer}}}, \bibnamefont{and}
  \bibinfo{author}{\bibfnamefont{G.}~\bibnamefont{{Mitselmakher}}},
  \bibinfo{journal}{Classical and Quantum Gravity}
  \textbf{\bibinfo{volume}{25}}, \bibinfo{eid}{114029} (\bibinfo{year}{2008}),
  \eprint{0802.3232}.

\bibitem[{\citenamefont{Lynch et~al.}(2017)\citenamefont{Lynch, Vitale, Essick,
  Katsavounidis, and Robinet}}]{PhysRevD.95.104046}
\bibinfo{author}{\bibfnamefont{R.}~\bibnamefont{Lynch}},
  \bibinfo{author}{\bibfnamefont{S.}~\bibnamefont{Vitale}},
  \bibinfo{author}{\bibfnamefont{R.}~\bibnamefont{Essick}},
  \bibinfo{author}{\bibfnamefont{E.}~\bibnamefont{Katsavounidis}},
  \bibnamefont{and} \bibinfo{author}{\bibfnamefont{F.}~\bibnamefont{Robinet}},
  \bibinfo{journal}{Phys. Rev. D} \textbf{\bibinfo{volume}{95}},
  \bibinfo{pages}{104046} (\bibinfo{year}{2017}),
  \urlprefix\url{https://link.aps.org/doi/10.1103/PhysRevD.95.104046}.

\bibitem[{\citenamefont{McIver}(2015)}]{McIver:2015pms}
\bibinfo{author}{\bibfnamefont{J.~L.} \bibnamefont{McIver}}, Ph.D. thesis,
  \bibinfo{school}{Massachusetts U., Amherst} (\bibinfo{year}{2015}),
  \urlprefix\url{https://scholarworks.umass.edu/dissertations_2/539/}.

\bibitem[{\citenamefont{{Heng}}(2009)}]{heng:09}
\bibinfo{author}{\bibfnamefont{I.~S.} \bibnamefont{{Heng}}},
  \bibinfo{journal}{\cqg} \textbf{\bibinfo{volume}{26}}, \bibinfo{eid}{105005}
  (\bibinfo{year}{2009}), \eprint{0810.5707}.

\bibitem[{\citenamefont{{R{\"o}ver} et~al.}(2009)\citenamefont{{R{\"o}ver},
  {Bizouard}, {Christensen}, {Dimmelmeier}, {Heng}, and
  {Meyer}}}]{2009PhRvD..80j2004R}
\bibinfo{author}{\bibfnamefont{C.}~\bibnamefont{{R{\"o}ver}}},
  \bibinfo{author}{\bibfnamefont{M.-A.} \bibnamefont{{Bizouard}}},
  \bibinfo{author}{\bibfnamefont{N.}~\bibnamefont{{Christensen}}},
  \bibinfo{author}{\bibfnamefont{H.}~\bibnamefont{{Dimmelmeier}}},
  \bibinfo{author}{\bibfnamefont{I.~S.} \bibnamefont{{Heng}}},
  \bibnamefont{and} \bibinfo{author}{\bibfnamefont{R.}~\bibnamefont{{Meyer}}},
  \bibinfo{journal}{\prd} \textbf{\bibinfo{volume}{80}}, \bibinfo{eid}{102004}
  (\bibinfo{year}{2009}), \eprint{0909.1093}.

\bibitem[{\citenamefont{{Dimmelmeier} et~al.}(2008)\citenamefont{{Dimmelmeier},
  {Ott}, {Marek}, and {Janka}}}]{dimmelmeier:08}
\bibinfo{author}{\bibfnamefont{H.}~\bibnamefont{{Dimmelmeier}}},
  \bibinfo{author}{\bibfnamefont{C.~D.} \bibnamefont{{Ott}}},
  \bibinfo{author}{\bibfnamefont{A.}~\bibnamefont{{Marek}}}, \bibnamefont{and}
  \bibinfo{author}{\bibfnamefont{H.-T.} \bibnamefont{{Janka}}},
  \bibinfo{journal}{\prd} \textbf{\bibinfo{volume}{78}},
  \bibinfo{pages}{064056} (\bibinfo{year}{2008}).

\bibitem[{\citenamefont{{Logue} et~al.}(2012)\citenamefont{{Logue}, {Ott},
  {Heng}, {Kalmus}, and {Scargill}}}]{logue:12}
\bibinfo{author}{\bibfnamefont{J.}~\bibnamefont{{Logue}}},
  \bibinfo{author}{\bibfnamefont{C.~D.} \bibnamefont{{Ott}}},
  \bibinfo{author}{\bibfnamefont{I.~S.} \bibnamefont{{Heng}}},
  \bibinfo{author}{\bibfnamefont{P.}~\bibnamefont{{Kalmus}}}, \bibnamefont{and}
  \bibinfo{author}{\bibfnamefont{J.}~\bibnamefont{{Scargill}}},
  \bibinfo{journal}{\prd} \textbf{\bibinfo{volume}{86}}, \bibinfo{eid}{044023}
  (\bibinfo{year}{2012}).

\bibitem[{\citenamefont{{Powell} et~al.}(2016)\citenamefont{{Powell}, {Gossan},
  {Logue}, and {Heng}}}]{powell:16b}
\bibinfo{author}{\bibfnamefont{J.}~\bibnamefont{{Powell}}},
  \bibinfo{author}{\bibfnamefont{S.~E.} \bibnamefont{{Gossan}}},
  \bibinfo{author}{\bibfnamefont{J.}~\bibnamefont{{Logue}}}, \bibnamefont{and}
  \bibinfo{author}{\bibfnamefont{I.~S.} \bibnamefont{{Heng}}},
  \bibinfo{journal}{\prd} \textbf{\bibinfo{volume}{94}}, \bibinfo{eid}{123012}
  (\bibinfo{year}{2016}), \eprint{1610.05573}.

\bibitem[{\citenamefont{{Powell}
  et~al.}(2017{\natexlab{a}})\citenamefont{{Powell}, {Szczepanczyk}, and
  {Heng}}}]{2017PhRvD..96l3013P}
\bibinfo{author}{\bibfnamefont{J.}~\bibnamefont{{Powell}}},
  \bibinfo{author}{\bibfnamefont{M.}~\bibnamefont{{Szczepanczyk}}},
  \bibnamefont{and} \bibinfo{author}{\bibfnamefont{I.~S.}
  \bibnamefont{{Heng}}}, \bibinfo{journal}{\prd} \textbf{\bibinfo{volume}{96}},
  \bibinfo{eid}{123013} (\bibinfo{year}{2017}{\natexlab{a}}),
  \eprint{1709.00955}.

\bibitem[{\citenamefont{Myers et~al.}(1980)\citenamefont{Myers, Rabiner, and
  Rosenberg}}]{myers1980performance}
\bibinfo{author}{\bibfnamefont{C.}~\bibnamefont{Myers}},
  \bibinfo{author}{\bibfnamefont{L.}~\bibnamefont{Rabiner}}, \bibnamefont{and}
  \bibinfo{author}{\bibfnamefont{A.}~\bibnamefont{Rosenberg}},
  \bibinfo{journal}{IEEE Transactions on Acoustics, Speech, and Signal
  Processing} \textbf{\bibinfo{volume}{28}}, \bibinfo{pages}{623}
  (\bibinfo{year}{1980}).

\bibitem[{\citenamefont{Myers and Rabiner}(1981)}]{myers1981comparative}
\bibinfo{author}{\bibfnamefont{C.~S.} \bibnamefont{Myers}} \bibnamefont{and}
  \bibinfo{author}{\bibfnamefont{L.~R.} \bibnamefont{Rabiner}},
  \bibinfo{journal}{Bell System Technical Journal}
  \textbf{\bibinfo{volume}{60}}, \bibinfo{pages}{1389} (\bibinfo{year}{1981}).

\bibitem[{\citenamefont{M{\"u}ller}(2007)}]{muller2007dynamic}
\bibinfo{author}{\bibfnamefont{M.}~\bibnamefont{M{\"u}ller}},
  \bibinfo{journal}{Information retrieval for music and motion} pp.
  \bibinfo{pages}{69--84} (\bibinfo{year}{2007}).

\bibitem[{\citenamefont{Sempena et~al.}(2011)\citenamefont{Sempena, Maulidevi,
  and Aryan}}]{sempena2011human}
\bibinfo{author}{\bibfnamefont{S.}~\bibnamefont{Sempena}},
  \bibinfo{author}{\bibfnamefont{N.~U.} \bibnamefont{Maulidevi}},
  \bibnamefont{and} \bibinfo{author}{\bibfnamefont{P.~R.} \bibnamefont{Aryan}},
  in \emph{\bibinfo{booktitle}{Electrical Engineering and Informatics (ICEEI),
  2011 International Conference on}} (\bibinfo{organization}{IEEE},
  \bibinfo{year}{2011}), pp. \bibinfo{pages}{1--5}.

\bibitem[{\citenamefont{ten Holt et~al.}(2007)\citenamefont{ten Holt, Reinders,
  and Hendriks}}]{ten2007multi}
\bibinfo{author}{\bibfnamefont{G.~A.} \bibnamefont{ten Holt}},
  \bibinfo{author}{\bibfnamefont{M.~J.} \bibnamefont{Reinders}},
  \bibnamefont{and} \bibinfo{author}{\bibfnamefont{E.}~\bibnamefont{Hendriks}},
  in \emph{\bibinfo{booktitle}{Thirteenth annual conference of the Advanced
  School for Computing and Imaging}} (\bibinfo{year}{2007}), vol.
  \bibinfo{volume}{300}, p.~\bibinfo{pages}{1}.

\bibitem[{\citenamefont{Martens and Claesen}(1996)}]{martens1996line}
\bibinfo{author}{\bibfnamefont{R.}~\bibnamefont{Martens}} \bibnamefont{and}
  \bibinfo{author}{\bibfnamefont{L.}~\bibnamefont{Claesen}}, in
  \emph{\bibinfo{booktitle}{Pattern Recognition, 1996., Proceedings of the 13th
  International Conference on}} (\bibinfo{organization}{IEEE},
  \bibinfo{year}{1996}), vol.~\bibinfo{volume}{3}, pp. \bibinfo{pages}{38--42}.

\bibitem[{\citenamefont{Bartolini et~al.}(2005)\citenamefont{Bartolini,
  Ciaccia, and Patella}}]{bartolini2005warp}
\bibinfo{author}{\bibfnamefont{I.}~\bibnamefont{Bartolini}},
  \bibinfo{author}{\bibfnamefont{P.}~\bibnamefont{Ciaccia}}, \bibnamefont{and}
  \bibinfo{author}{\bibfnamefont{M.}~\bibnamefont{Patella}},
  \bibinfo{journal}{IEEE transactions on pattern analysis and machine
  intelligence} \textbf{\bibinfo{volume}{27}}, \bibinfo{pages}{142}
  (\bibinfo{year}{2005}).

\bibitem[{\citenamefont{Scheidegger et~al.}(2008)\citenamefont{Scheidegger,
  Fischer, Whitehouse, and Liebend{\"o}rfer}}]{scheidegger2008gravitational}
\bibinfo{author}{\bibfnamefont{S.}~\bibnamefont{Scheidegger}},
  \bibinfo{author}{\bibfnamefont{T.}~\bibnamefont{Fischer}},
  \bibinfo{author}{\bibfnamefont{S.}~\bibnamefont{Whitehouse}},
  \bibnamefont{and}
  \bibinfo{author}{\bibfnamefont{M.}~\bibnamefont{Liebend{\"o}rfer}},
  \bibinfo{journal}{Astronomy \& astrophysics} \textbf{\bibinfo{volume}{490}},
  \bibinfo{pages}{231} (\bibinfo{year}{2008}).

\bibitem[{\citenamefont{{Janka}}(2012)}]{2012ARNPS..62..407J}
\bibinfo{author}{\bibfnamefont{H.-T.} \bibnamefont{{Janka}}},
  \bibinfo{journal}{Annual Review of Nuclear and Particle Science}
  \textbf{\bibinfo{volume}{62}}, \bibinfo{pages}{407} (\bibinfo{year}{2012}),
  \eprint{1206.2503}.

\bibitem[{\citenamefont{{Andresen} et~al.}(2018)\citenamefont{{Andresen},
  {M{\"u}ller}, {Janka}, {Summa}, {Gill}, and {Zanolin}}}]{2018arXiv181007638A}
\bibinfo{author}{\bibfnamefont{H.}~\bibnamefont{{Andresen}}},
  \bibinfo{author}{\bibfnamefont{E.}~\bibnamefont{{M{\"u}ller}}},
  \bibinfo{author}{\bibfnamefont{H.-T.} \bibnamefont{{Janka}}},
  \bibinfo{author}{\bibfnamefont{A.}~\bibnamefont{{Summa}}},
  \bibinfo{author}{\bibfnamefont{K.}~\bibnamefont{{Gill}}}, \bibnamefont{and}
  \bibinfo{author}{\bibfnamefont{M.}~\bibnamefont{{Zanolin}}},
  \bibinfo{journal}{ArXiv e-prints}  (\bibinfo{year}{2018}),
  \eprint{1810.07638}.

\bibitem[{\citenamefont{{O'Connor} and {Couch}}(2018)}]{oconnor_18}
\bibinfo{author}{\bibfnamefont{E.~P.} \bibnamefont{{O'Connor}}}
  \bibnamefont{and} \bibinfo{author}{\bibfnamefont{S.~M.}
  \bibnamefont{{Couch}}}, \bibinfo{journal}{\apj}
  \textbf{\bibinfo{volume}{865}}, \bibinfo{eid}{81} (\bibinfo{year}{2018}),
  \eprint{1807.07579}.

\bibitem[{\citenamefont{Blondin et~al.}(2003)\citenamefont{Blondin, Mezzacappa,
  and DeMarino}}]{0004-637X-584-2-971}
\bibinfo{author}{\bibfnamefont{J.~M.} \bibnamefont{Blondin}},
  \bibinfo{author}{\bibfnamefont{A.}~\bibnamefont{Mezzacappa}},
  \bibnamefont{and} \bibinfo{author}{\bibfnamefont{C.}~\bibnamefont{DeMarino}},
  \bibinfo{journal}{The Astrophysical Journal} \textbf{\bibinfo{volume}{584}},
  \bibinfo{pages}{971} (\bibinfo{year}{2003}),
  \urlprefix\url{http://stacks.iop.org/0004-637X/584/i=2/a=971}.

\bibitem[{\citenamefont{{Blondin} and
  {Mezzacappa}}(2006)}]{2006ApJ...642..401B}
\bibinfo{author}{\bibfnamefont{J.~M.} \bibnamefont{{Blondin}}}
  \bibnamefont{and}
  \bibinfo{author}{\bibfnamefont{A.}~\bibnamefont{{Mezzacappa}}},
  \bibinfo{journal}{\apj} \textbf{\bibinfo{volume}{642}}, \bibinfo{pages}{401}
  (\bibinfo{year}{2006}), \eprint{astro-ph/0507181}.

\bibitem[{\citenamefont{{Foglizzo} et~al.}(2007)\citenamefont{{Foglizzo},
  {Galletti}, {Scheck}, and {Janka}}}]{2007ApJ...654.1006F}
\bibinfo{author}{\bibfnamefont{T.}~\bibnamefont{{Foglizzo}}},
  \bibinfo{author}{\bibfnamefont{P.}~\bibnamefont{{Galletti}}},
  \bibinfo{author}{\bibfnamefont{L.}~\bibnamefont{{Scheck}}}, \bibnamefont{and}
  \bibinfo{author}{\bibfnamefont{H.-T.} \bibnamefont{{Janka}}},
  \bibinfo{journal}{\apj} \textbf{\bibinfo{volume}{654}}, \bibinfo{pages}{1006}
  (\bibinfo{year}{2007}), \eprint{astro-ph/0606640}.

\bibitem[{\citenamefont{{Beck} et~al.}(2012)\citenamefont{{Beck}, {Montalban},
  {Kallinger}, {De Ridder}, {Aerts}, {Garc{\'{\i}}a}, {Hekker}, {Dupret},
  {Mosser}, {Eggenberger} et~al.}}]{2012Natur.481...55B}
\bibinfo{author}{\bibfnamefont{P.~G.} \bibnamefont{{Beck}}},
  \bibinfo{author}{\bibfnamefont{J.}~\bibnamefont{{Montalban}}},
  \bibinfo{author}{\bibfnamefont{T.}~\bibnamefont{{Kallinger}}},
  \bibinfo{author}{\bibfnamefont{J.}~\bibnamefont{{De Ridder}}},
  \bibinfo{author}{\bibfnamefont{C.}~\bibnamefont{{Aerts}}},
  \bibinfo{author}{\bibfnamefont{R.~A.} \bibnamefont{{Garc{\'{\i}}a}}},
  \bibinfo{author}{\bibfnamefont{S.}~\bibnamefont{{Hekker}}},
  \bibinfo{author}{\bibfnamefont{M.-A.} \bibnamefont{{Dupret}}},
  \bibinfo{author}{\bibfnamefont{B.}~\bibnamefont{{Mosser}}},
  \bibinfo{author}{\bibfnamefont{P.}~\bibnamefont{{Eggenberger}}},
  \bibnamefont{et~al.}, \bibinfo{journal}{\nat} \textbf{\bibinfo{volume}{481}},
  \bibinfo{pages}{55} (\bibinfo{year}{2012}), \eprint{1112.2825}.

\bibitem[{\citenamefont{{Mosser} et~al.}(2012)\citenamefont{{Mosser}, {Goupil},
  {Belkacem}, {Marques}, {Beck}, {Bloemen}, {De Ridder}, {Barban}, {Deheuvels},
  {Elsworth} et~al.}}]{2012A&A...548A..10M}
\bibinfo{author}{\bibfnamefont{B.}~\bibnamefont{{Mosser}}},
  \bibinfo{author}{\bibfnamefont{M.~J.} \bibnamefont{{Goupil}}},
  \bibinfo{author}{\bibfnamefont{K.}~\bibnamefont{{Belkacem}}},
  \bibinfo{author}{\bibfnamefont{J.~P.} \bibnamefont{{Marques}}},
  \bibinfo{author}{\bibfnamefont{P.~G.} \bibnamefont{{Beck}}},
  \bibinfo{author}{\bibfnamefont{S.}~\bibnamefont{{Bloemen}}},
  \bibinfo{author}{\bibfnamefont{J.}~\bibnamefont{{De Ridder}}},
  \bibinfo{author}{\bibfnamefont{C.}~\bibnamefont{{Barban}}},
  \bibinfo{author}{\bibfnamefont{S.}~\bibnamefont{{Deheuvels}}},
  \bibinfo{author}{\bibfnamefont{Y.}~\bibnamefont{{Elsworth}}},
  \bibnamefont{et~al.}, \bibinfo{journal}{\aap} \textbf{\bibinfo{volume}{548}},
  \bibinfo{eid}{A10} (\bibinfo{year}{2012}), \eprint{1209.3336}.

\bibitem[{\citenamefont{{Scheidegger} et~al.}(2010)\citenamefont{{Scheidegger},
  {K{\"a}ppeli}, {Whitehouse}, {Fischer}, and
  {Liebend{\"o}rfer}}}]{scheidegger:10b}
\bibinfo{author}{\bibfnamefont{S.}~\bibnamefont{{Scheidegger}}},
  \bibinfo{author}{\bibfnamefont{R.}~\bibnamefont{{K{\"a}ppeli}}},
  \bibinfo{author}{\bibfnamefont{S.~C.} \bibnamefont{{Whitehouse}}},
  \bibinfo{author}{\bibfnamefont{T.}~\bibnamefont{{Fischer}}},
  \bibnamefont{and}
  \bibinfo{author}{\bibfnamefont{M.}~\bibnamefont{{Liebend{\"o}rfer}}},
  \bibinfo{journal}{\aap} \textbf{\bibinfo{volume}{514}}, \bibinfo{pages}{A51}
  (\bibinfo{year}{2010}).

\bibitem[{\citenamefont{Jolliffe}(2011)}]{jolliffe2011principal}
\bibinfo{author}{\bibfnamefont{I.}~\bibnamefont{Jolliffe}}, in
  \emph{\bibinfo{booktitle}{International encyclopedia of statistical science}}
  (\bibinfo{publisher}{Springer}, \bibinfo{year}{2011}), pp.
  \bibinfo{pages}{1094--1096}.

\bibitem[{\citenamefont{Peres-Neto et~al.}(2005)\citenamefont{Peres-Neto,
  Jackson, and Somers}}]{peres2005many}
\bibinfo{author}{\bibfnamefont{P.~R.} \bibnamefont{Peres-Neto}},
  \bibinfo{author}{\bibfnamefont{D.~A.} \bibnamefont{Jackson}},
  \bibnamefont{and} \bibinfo{author}{\bibfnamefont{K.~M.}
  \bibnamefont{Somers}}, \bibinfo{journal}{Computational Statistics \& Data
  Analysis} \textbf{\bibinfo{volume}{49}}, \bibinfo{pages}{974}
  (\bibinfo{year}{2005}).

\bibitem[{\citenamefont{Karhunen et~al.}(1998)\citenamefont{Karhunen, Pajunen,
  and Oja}}]{karhunen1998nonlinear}
\bibinfo{author}{\bibfnamefont{J.}~\bibnamefont{Karhunen}},
  \bibinfo{author}{\bibfnamefont{P.}~\bibnamefont{Pajunen}}, \bibnamefont{and}
  \bibinfo{author}{\bibfnamefont{E.}~\bibnamefont{Oja}},
  \bibinfo{journal}{Neurocomputing} \textbf{\bibinfo{volume}{22}},
  \bibinfo{pages}{5} (\bibinfo{year}{1998}).

\bibitem[{\citenamefont{Sakoe and Chiba}(1978)}]{sakoe1978dynamic}
\bibinfo{author}{\bibfnamefont{H.}~\bibnamefont{Sakoe}} \bibnamefont{and}
  \bibinfo{author}{\bibfnamefont{S.}~\bibnamefont{Chiba}},
  \bibinfo{journal}{IEEE transactions on acoustics, speech, and signal
  processing} \textbf{\bibinfo{volume}{26}}, \bibinfo{pages}{43}
  (\bibinfo{year}{1978}).

\bibitem[{\citenamefont{{Abbott} et~al.}(2018)\citenamefont{{Abbott}, {Abbott},
  {Abbott}, {Abernathy}, {Acernese}, {Ackley}, {Adams}, {Adams}, {Addesso},
  {Adhikari} et~al.}}]{2018LRR....21....3A}
\bibinfo{author}{\bibfnamefont{B.~P.} \bibnamefont{{Abbott}}},
  \bibinfo{author}{\bibfnamefont{R.}~\bibnamefont{{Abbott}}},
  \bibinfo{author}{\bibfnamefont{T.~D.} \bibnamefont{{Abbott}}},
  \bibinfo{author}{\bibfnamefont{M.~R.} \bibnamefont{{Abernathy}}},
  \bibinfo{author}{\bibfnamefont{F.}~\bibnamefont{{Acernese}}},
  \bibinfo{author}{\bibfnamefont{K.}~\bibnamefont{{Ackley}}},
  \bibinfo{author}{\bibfnamefont{C.}~\bibnamefont{{Adams}}},
  \bibinfo{author}{\bibfnamefont{T.}~\bibnamefont{{Adams}}},
  \bibinfo{author}{\bibfnamefont{P.}~\bibnamefont{{Addesso}}},
  \bibinfo{author}{\bibfnamefont{R.~X.} \bibnamefont{{Adhikari}}},
  \bibnamefont{et~al.}, \bibinfo{journal}{Living Reviews in Relativity}
  \textbf{\bibinfo{volume}{21}}, \bibinfo{eid}{3} (\bibinfo{year}{2018}),
  \eprint{1304.0670}.

\bibitem[{\citenamefont{{Powell} et~al.}(2015)\citenamefont{{Powell},
  {Trifir{\`o}}, {Cuoco}, {Heng}, and {Cavagli{\`a}}}}]{2015CQGra..32u5012P}
\bibinfo{author}{\bibfnamefont{J.}~\bibnamefont{{Powell}}},
  \bibinfo{author}{\bibfnamefont{D.}~\bibnamefont{{Trifir{\`o}}}},
  \bibinfo{author}{\bibfnamefont{E.}~\bibnamefont{{Cuoco}}},
  \bibinfo{author}{\bibfnamefont{I.~S.} \bibnamefont{{Heng}}},
  \bibnamefont{and}
  \bibinfo{author}{\bibfnamefont{M.}~\bibnamefont{{Cavagli{\`a}}}},
  \bibinfo{journal}{Classical and Quantum Gravity}
  \textbf{\bibinfo{volume}{32}}, \bibinfo{eid}{215012} (\bibinfo{year}{2015}),
  \eprint{1505.01299}.

\bibitem[{\citenamefont{{Powell}
  et~al.}(2017{\natexlab{b}})\citenamefont{{Powell}, {Torres-Forn{\'e}},
  {Lynch}, {Trifir{\`o}}, {Cuoco}, {Cavagli{\`a}}, {Heng}, and
  {Font}}}]{2017CQGra..34c4002P}
\bibinfo{author}{\bibfnamefont{J.}~\bibnamefont{{Powell}}},
  \bibinfo{author}{\bibfnamefont{A.}~\bibnamefont{{Torres-Forn{\'e}}}},
  \bibinfo{author}{\bibfnamefont{R.}~\bibnamefont{{Lynch}}},
  \bibinfo{author}{\bibfnamefont{D.}~\bibnamefont{{Trifir{\`o}}}},
  \bibinfo{author}{\bibfnamefont{E.}~\bibnamefont{{Cuoco}}},
  \bibinfo{author}{\bibfnamefont{M.}~\bibnamefont{{Cavagli{\`a}}}},
  \bibinfo{author}{\bibfnamefont{I.~S.} \bibnamefont{{Heng}}},
  \bibnamefont{and} \bibinfo{author}{\bibfnamefont{J.~A.}
  \bibnamefont{{Font}}}, \bibinfo{journal}{Classical and Quantum Gravity}
  \textbf{\bibinfo{volume}{34}}, \bibinfo{eid}{034002}
  (\bibinfo{year}{2017}{\natexlab{b}}), \eprint{1609.06262}.

\end{thebibliography}

\end{document}